\documentclass[oneside,english]{amsart}

\usepackage[T1]{fontenc}
\usepackage[latin9]{inputenc}
\usepackage[a4paper]{geometry}
\usepackage{amsthm}

\makeatletter
\numberwithin{equation}{section}
\numberwithin{figure}{section}
\theoremstyle{plain}
\newtheorem{thm}{\protect\theoremname}
  \theoremstyle{plain}
  \newtheorem{lem}[thm]{\protect\lemmaname}

\makeatother

\usepackage{babel}
  \providecommand{\lemmaname}{Lemma}
\providecommand{\theoremname}{Theorem}

\begin{document}

\title[Matthew Radley Brown 28 June 2015]{the strength of a pair of point vortices in an incompressible inviscid
fluid in 3d can blow up in finite time}

\author{Matthew Radley Brown}

\date{28 June 2015, mbro009@aucklanduni.ac.nz}

\email{mbro009@aucklanduni.ac.nz}
\begin{abstract}
The evolution of a pair of point vortices in $R^{3}$, subject to
the inviscid Euler equations for incompressible fluid flow, is solved
exactly for rotationally symmetric initial conditions. This exact
solution shows that the vortex strength for a pair of point vortices
can either remain stable or blow up in finite time, depending on the
initial data.
\end{abstract}
\maketitle

\section{Introduction}

The evolution of the vortex strengths and orientations of a pair of
point vortices in an inviscid incompressible 3d fluid can be modelled
\cite{MB} using the following dynamical equations
for the vortex strengths and orientations: 
\begin{eqnarray}
\omega_{a}^{\prime}\left(t\right) & = & \frac{-3}{8\pi}\left[\frac{K\left(x_{a}-x_{b},\omega_{b}\right)}{\left|x_{a}-x_{b}\right|^{5}}\right]\omega_{a},\label{eq:vort}\\
\omega_{b}^{\prime}\left(t\right) & = & \frac{-3}{8\pi}\left[\frac{K\left(x_{b}-x_{a},\omega_{a}\right)}{\left|x_{b}-x_{a}\right|^{5}}\right]\omega_{b},\label{eq:vort2}
\end{eqnarray}
and the following equations for the locations of the point vortices:
\begin{eqnarray}
x_{a}^{\prime}\left(t\right) & = & \frac{1}{4\pi}\frac{\omega_{b}\times\left(x_{a}-x_{b}\right)}{\left|x_{a}-x_{b}\right|^{3}},\label{eq:pos1}\\
x_{b}^{\prime}\left(t\right) & = & \frac{1}{4\pi}\frac{\omega_{a}\times\left(x_{b}-x_{a}\right)}{\left|x_{b}-x_{a}\right|^{3}},\label{eq:pos2}
\end{eqnarray}
and the matrix $K\left(x,\omega\right)=\left[\omega\times x\right]\otimes x+x\otimes\left[\omega\times x\right]$
. These equations are solved below for a pair of point vortices with
rotationally symmetric initial conditions. The exact solution shows
that the vortex strength can either remain stable or blow up in finite
time. Note that the solution blows up in spite of the fact that the
point vortex cores do not approach each other.

\subsection{Eigenvalues and eigenvectors of $K\left(x,\omega\right)$ }

The 3x3 matrix $K\left(x,\omega\right)$ is real and symmetric so
has real eigenvalues and an orthonormal basis of eigenvectors. The
eigenvalues of $K\left(x,\omega\right)$ are $0$,$\pm\lambda$ where
$\lambda=\left|x\times\omega\right|\left|x\right|$. Thus, transformation
of any vector by the matrix $K\left(x,\omega\right)$ is equivalent
to (1) a projection onto the plane co-planar with $e_{\pm}$ followed
by (2) a reflection in the plane co-planar with $e_{0}$ and $e_{+}$
followed by (3) a stretch of magnitude $\lambda$. The eigenvector
$e_{0}$ corresponding to the eigenvalue $\lambda=0$ is $e_{0}=x\times\left(\omega\times x\right)$.
That is, both normal to $x$ and co-planar with the plane formed by
$x$ and $\omega$. The eigenvectors for the eigenvalues $\pm\left|x\times\omega\right|\left|x\right|$
are $e_{\pm}=R_{e_{0},\pm\frac{\pi}{4}}x$ . That is, they correspond
to the vector $x$ rotated both $\pm\frac{\pi}{4}$ around the axis
defined by $e_{0}$. 

These show that it is possible for the first point vortex to augment
the growth in magnitude of the vorticity strength of the second point
vortex, while itself being augmented by the second.

\section{Exact Solution for a pair of point vortices for rotationally symmetric
initial conditions}

The exact solution presented below consists of a pair of rotationally
symmetric point vortices rotating around the axis defined by $\left(0,0,1\right)^{T}$.
Note the magnitude of the vorticity increases more rapidly than the
location of the point vortices varies. 
\begin{lem}
Let the initial data comprise two point vortices located at:\textrm{\textup{
\[
x_{a}\left(0\right)=\left(\begin{array}{ccc}
-\frac{1}{2} & 0 & 0\end{array}\right)^{T},x_{b}\left(0\right)=\left(\begin{array}{ccc}
\frac{1}{2} & 0 & 0\end{array}\right)^{T}
\]
}}
\end{lem}
with vortex strength: 
\[
\omega_{a}\left(0\right)=\left(\begin{array}{ccc}
-\sqrt{\frac{1}{6}} & -\sqrt{\frac{7}{12}} & \frac{1}{2}\end{array}\right)^{T},\omega_{b}\left(0\right)=\left(\begin{array}{ccc}
\sqrt{\frac{1}{6}} & \sqrt{\frac{7}{12}} & \frac{1}{2}\end{array}\right)^{T}.
\]
respectively. Then the solution to equations \ref{eq:vort}-\ref{eq:pos2}
for this initial data is: 
\begin{eqnarray*}
\left[x_{a}\left(t\right);x_{b}\left(t\right)\right] & = & R_{k,\phi}\left(t\right)\left[\begin{array}{c}
-\frac{1}{2}\\
0\\
z\left(t\right)
\end{array};\begin{array}{c}
\frac{1}{2}\\
0\\
z\left(t\right)
\end{array}\right]
\end{eqnarray*}
and
\begin{eqnarray*}
\left[\omega_{a}\left(t\right);\omega_{b}\left(t\right)\right] & = & \omega\left(t\right)R_{k,\phi}\left(t\right)\left[\omega_{a}\left(0\right);\omega_{b}\left(0\right)\right]
\end{eqnarray*}
where:
\[
R_{k,\phi}\left(t\right)=\left[\begin{array}{ccc}
\cos\phi\left(t\right) & -\sin\phi\left(t\right) & 0\\
\sin\phi\left(t\right) & \cos\phi\left(t\right) & 0\\
0 & 0 & 1
\end{array}\right]
\]
and
\begin{eqnarray*}
z\left(t\right) & = & \sqrt{\frac{2}{3}}\log\left(1-\frac{1}{8\pi}\sqrt{\frac{7}{2}}t\right)\\
\phi\left(t\right) & = & 2\sqrt{\frac{2}{7}}\log\left(1-\frac{1}{8\pi}\sqrt{\frac{7}{2}}t\right)\\
\omega\left(t\right) & = & \frac{1}{\frac{1}{8\pi}\sqrt{\frac{7}{2}}t-1}.
\end{eqnarray*}
which blows up after time $t^{*}=8\pi\sqrt{\frac{2}{7}}$.
\begin{proof}
Substituting into the equations of motion verifies the solution.
\end{proof}
If the orientation of the initial conditions are reversed
then the vorticity strength diminishes to $0$ instead of blowing
up.

\end{document}